# Schrödinger's Red Pixel by Quasi Bound-State-In-Continuum

Zhaogang Dong[1,†,*], Lei Jin[2,3,†], Soroosh Daqiqeh Rezaei[4], Hao Wang[4], Yang Chen[2], Febiana Tjiptoharsono[1], Jinfa Ho[1], Sergey Gorelik,[5] Ray Jia Hong Ng[4], Qifeng Ruan[4], Cheng-Wei Qiu[2,*], and Joel K. W. Yang[1,4,*]

[1]Institute of Materials Research and Engineering, A*STAR (Agency for Science, Technology and Research), 2 Fusionopolis Way, #08-03 Innovis, 138634 Singapore

[2]Department of Electrical and Computer Engineering, National University of Singapore, 4 Engineering Drive 3, Singapore 117583, Singapore

[3]College of Electronic and Information Engineering, Hangzhou Dianzi University, Hangzhou, China, 310018

[4]Singapore University of Technology and Design, 8 Somapah Road, 487372, Singapore

[5]Singapore Institute of Food and Biotechnology Innovation, A*STAR (Agency for Science, Technology and Research), 31 Biopolis Way, #01-02 Nanos, Singapore 138669

*Correspondence and requests for materials should be addressed to J.K.W.Y. (email: joel_yang@sutd.edu.sg), C.-W.Q. (email: chengwei.qiu@nus.edu.sg), and Z.D. (email: dongz@imre.a-star.edu.sg).

**Abstract**

While structural colors are ubiquitous in nature, saturated reds are mysteriously absent. Hence, a longstanding problem is in fabricating nanostructured surfaces that exhibit reflectance approaching the theoretical limit. This limit is termed the Schrödinger red and demands sharp spectral transitions from "stopband" to a high reflectance "passband" with total suppression of higher-order resonances at blue and green wavelengths. Current approaches based on metallic or dielectric nanoantennas are insufficient to simultaneously meet these conditions. Here, for the 1st time, we designed and fabricated tall Si nanoantenna arrays on quartz substrate to support two partially overlapping *y*-polarized quasi bound-state-in-the-continuum (q-BIC) modes in the red wavelengths with sharp spectral edges. These structures produce possibly the most saturated and brightest reds with ~80% reflectance, exceeding the red vertex in sRGB and even the cadmium red pigment. We employed a gradient descent algorithm with structures supporting q-BIC as the starting point. Although the current design is polarization dependent, the proposed paradigm has enabled us to achieve the elusive structural red and the design principle could be generalized to Schrödinger's pixels of other colors. The design is suitable for scale up using other nanofabrication techniques for larger area applications, such as red pixels in displays, decorative coatings, and miniaturized spectrometers with high wavelength selectivity.

Dye-free nanostructural colors offer unprecedented print resolution, fade resistance, scalable manufacturability and viewing angle independence.[1] High-resolution color printing beyond the optical diffraction limit was first demonstrated with metallic nanostructures with localized plasmon resonances, such as gold[2] and aluminum.[3, 4] However, the achieved color saturation of these plasmonic nanostructures is limited due in part to Ohmic losses in metals. On the other hand, with lower Ohmic losses, dielectric nanostructures support Mie resonances[5-19] that produce color pixels with improved saturations due in part to abrupt spectral transitions. For instance, Si nanostructures that behave as if suspended in in free-space when placed on $Si_3N_4$ anti-reflection layer achieve Kerker's conditions that sharpen spectral transitions for highly saturated colors.[15] Similarly, $TiO_2$ nanostructures with electric dipole and magnetic dipole resonances have demonstrated high color saturation[16] and tunability.[17] Furthermore, the color saturation has been improved further by the multi-layer dielectric stacked nanostructures[18] and refractive index matching layers.[20]

Despite significant progress in nanostructural colors,[2-4, 15-18, 20-26] a longstanding challenge still remains particularly in achieving highly saturated reds, with similar limitations observed in structural colors in nature.[27] For example, bird feathers exhibit only saturated structural blues and greens, but not reds.[27, 28] For instance, even the "red" feathers of parrots (Aves: Psittaciformes) have a magenta appearance as they exhibit significant blue and green components in the reflectance spectrum.[29] In addition, the scales of longhorn beetles[30] and natural cuttlefish ink[31] have structural colors spanning the visible spectrum, with the exception of saturated red (closest being magenta). This elusive-red phenomenon could be partially explained by the underlying physics: A highly saturated red pixel requires the fundamental structural resonance beyond 600 nm in wavelength. However, resonators that support fundamental modes in red would also support higher-order modes at shorter wavelengths (*i.e.*, 380 nm to 480 nm). Consequently, these high-order modes will shift the color towards blue in the chromaticity diagram, deteriorating its red saturation and eventually result in magenta instead.[32] In other words, generating a highly saturated red in reflection mode requires a high reflectance of the red wavelengths (600 nm and longer) and a concomitant near-total suppression of reflectance or scattering at other wavelengths. An ideal reflectance spectrum is one with unity reflectance in the desired wavelength range and zero reflectance elsewhere, which was recently coined the Schrödinger pixel based on Erwin Schrödinger's calculations.[26, 33] Incidentally, highly saturated reds are also mysteriously rare in synthetic and natural pigments and dyes based on light absorption,[27, 32] where these pigments and dyes are working under white light illumination, being different from emissive devices, such as light emitting diodes and lasers.

In this paper, we designed a Si nanoantenna array with two quasi bound-state-in-the-continuum (q-BIC) modes to achieve the ideal Schrödinger's red pixel with a high color saturation, by suppressing the high-order q-BIC modes at blue/green wavelengths *via* engineering the substrate-induced leakage effect and the intrinsic absorption of amorphous silicon. In simulation, the achieved color saturation in International Commission on Illumination (CIE) coordinate is (0.684, 0.301), which is much better than the red in the standard RGB (sRGB) triangle or the famous "cadmium red" pigment with the respective CIE coordinates of (0.64, 0.33) and (0.621, 0.32).[34] The geometry optimization of q-BIC antenna is based on a gradient descent approach integrated with finite-difference time-domain simulation, to search for the brightest and most saturated red *via* suppressing reflectance at blue and green wavelengths. Distinct from the more common *x*-polarized q-BIC modes in the literature,[35-37] here we introduce *y*-polarized q-BIC modes to achieve a better red color saturation and higher reflectance. Experimentally, the achieved red saturation indeed exceeds the sRGB triangle with a CIE coordinate of (0.654, 0.301) while maintaining a high reflectance of ~80%. The achieved highly saturated red could be potentially scaled up, e.g. *via* deep UV lithography[38], to reach the dimensions of reflective displays based on multi-layer film configuration[39-41] and lead to potential applications of compact red filters,[42] highly saturated reflective displays[26, 43, 44] and miniaturized spectrometers.[45-47]

**Design of BIC antenna array for achieving the ideal Schrödinger's red pixel.** Fig. 1a presents the reflectance spectrum of an ideal Schrödinger's red pixel, which has a flat top profile with a sharp transition from "stopband" to "passband". In comparison, Fig. 1a also presents the schematic illustration of a typical nanophotonic optical resonance, which cannot achieve highly saturated red due to the following two reasons. First, when the nanophotonic optical resonance has a fundamental mode at the red wavelength, but it usually has the higher order mode at the blue wavelength.[15, 16] Second, the broad resonances are significantly different from the Schrödinger's type resonance with steep edges.

On the other hand, bound-states-in-the-continuum (BIC) is an emerging concept in nanophotonics,[35-37, 48-55] and this BIC mode could potentially provide highly saturated reds. As the true BIC mode is optically inaccessible,[48] here we utilize the quasi-BIC (q-BIC) mode with broken symmetry.[35, 36, 48] Fig. 1b presents the simulated reflectance spectrum of the designed q-BIC antenna array with two BIC resonances at *y*-polarized incidence condition, where these *y*-polarized q-BIC resonances were not explored so far. Here, the merging of two q-BIC modes provides an effective approach to construct the ideal Schrödinger's red pixel due to two design principles. First, the high quality-factor of q-BIC mode will give rise to steep edges to enable sharp transition from "stopband" to "passband". Meanwhile, the merging of two q-BIC resonances will generate considerable full-width-at-half-maximum (FWHM), which is required for saturated red with high brightness. Although 1D photonic crystal is able to achieve a bandstop with a sharp transition too,[56] the blue color wavelength component was not fully suppressed and it has a limited printing resolution due to the multi-layer design with the infinite extension along *x*- and *y*-directions.

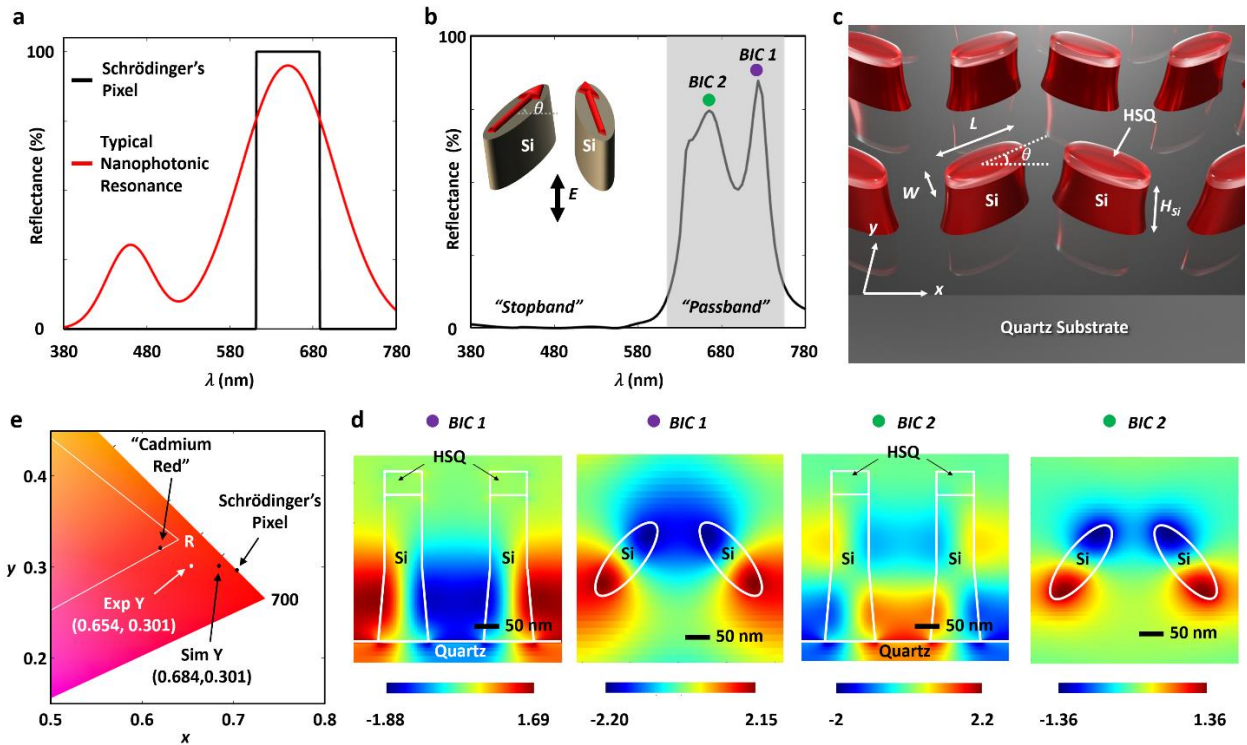

**Fig. 1 | Design principle, geometry, and results of BIC antenna array approaching the ideal Schrödinger's red pixel with high saturation. a,** Spectral profile comparison between the ideal Schrödinger's red pixel and a typical nanophotonic resonance. **b,** Simulated reflectance spectrum of the designed q-BIC antenna array with two q-BIC resonances to construct the ideal Schrödinger's red pixel. **c,** 3D schematic of showing the designed BIC antenna array based on amorphous silicon (a-Si) nanoantenna array. Each unit cell consists of a pair of elliptical a-Si nanostructures on a quartz substrate. **d**, Spatial distribution of $E_z$ at the two q-BIC resonant wavelengths. **e,** International Commission on Illumination (CIE) 1931 chromaticity diagram for

illustrating the optimization of red saturation of showing the achieved red CIE coordinate in simulation and experiment, in benchmarking to the famous red dye "cadmium red".

The simulated reflectance spectrum in Fig. 1b was obtained based a careful optimization of the q-BIC antenna design with the corresponding 3D schematic as shown in Fig. 1c. This q-BIC antenna is based on amorphous Si, with $n$ and $k$ values as shown in Supplementary Fig. 1. The designed unit cell consists of a pair of elliptical amorphous Si nanostructures with a long axis $L$ and a short axis $W$. Its long axis has an angle $\theta$ of with respect to $x$-axis. The pitch along $x$- and $y$-direction are denoted as $\Lambda_x$ and $\Lambda_y$, respectively. These parameters will be used in the optimization algorithm as discussed later. In addition, Fig. 1d presents the mode patterns of two q-BIC modes being excited on the antenna array. Moreover, Fig. 1e presents the achieved color saturation in a small section of the CIE 1931 chromaticity diagram, with the CIE coordinates of (0.684, 0.301) in simulation and (0.654, 0.301) in experiment, where the achieved red saturation is beyond the sRGB triangle. These achieved reds are more saturated than the "cadmium red" pigment, known for producing vibrant reds in modern art, with the CIE coordinate of (0.621, 0.320). In addition, based on the reflectance spectrum of Schrödinger's red pixel in Fig. 1a with a flat top profile from 612 nm to 688 nm, the CIE coordinate is (0.704, 0.296) as shown in Fig. 1e for benchmarking purpose. When the central wavelength of this Schrödinger's red pixel is shifting towards 700 nm, the coordinate will approach towards the CIE red apex, which is ultimate limit of red saturation.

Next, we present the gradient descent algorithm as integrated with FDTD simulation package to optimize the geometric parameters of the q-BIC antenna array. In Fig. 2a, our 1st cost function is the distance "$D$" on the CIE diagram from the achieved structural color coordinate to the red apex of the CIE plot, which is defined here as the reddest possible red corresponding to a monochromatic light source with a wavelength of 700 nm. Unfortunately, a surface with a narrow but perfect reflectance peak at 700 nm would result in a nearly black reflective appearance. A smaller *D* value corresponds to a more saturated red. In addition, we introduced a 2$^{nd}$ cost function to mitigate the tendency of the gradient descent algorithm to be stuck at a local minimum. The 2$^{nd}$ cost function maximizes the "CIE *x* coordinate", so that the optimization algorithm will find the optimized geometry towards achieving the CIE plot rim at the red corner. The cost functions will be toggled from one to the other when the optimization of one cost function is stuck at a local minimum.

The detailed algorithm flow is shown in Fig. 2b. Here, the cross sectional profile of the optimized Si antenna is based on the experimental result. For instance, a typical cross sectional profile of the fabricated Si nanostructures is shown in the Supplementary Fig. 2, where the etched Si nanostructures do not have perfectly vertical cross sectional profile, due to dry etching conditions. Instead, we observe two segments. The top ~154 nm segment (*i.e.*, the part close to the HSQ resist) has as ~90° side wall profile, while the bottom segment tapers outwards to a trapezoidal shape with side wall angle of ~82°. The modeled antenna cross section is shown in Supplementary Fig. 3.

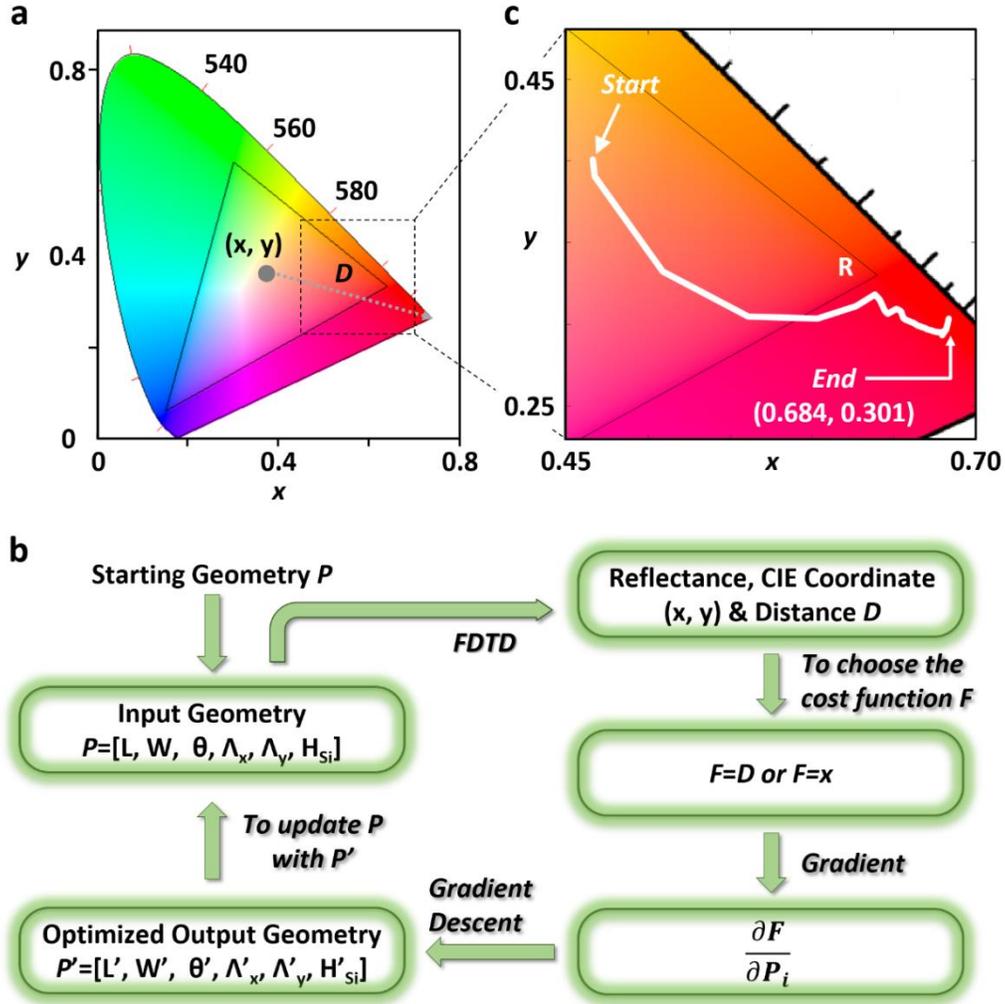

**Fig. 2 | Gradient descent optimization algorithm for optimizing the designed q-BIC antenna array to achieve the highly saturated red. a,** International Commission on Illumination (CIE) 1931 chromaticity diagram for illustrating the optimization of red saturation. The CIE coordinate of the simulated color pixel is denoted as ($x$, $y$). $D$ defines the distance between the simulated CIE coordinate ($x$, $y$) and the apex of highly saturated red at 700 nm, in the chromaticity diagram. **b,** Multi-parameter optimization based on the gradient descent algorithm as integrated with finite-difference time-domain (FDTD) simulation. Here, we define two cost functions that are toggled from one to another to allow escape from local minima. The 1$^{st}$ cost function is to minimize the distance value $D$, and the 2$^{nd}$ cost function is to maximize the CIE coordinate $x$ (see the main text

for the detailed optimization process). **c,** Trajectory of the CIE coordinates during the optimization process from the "starting point" to the "end point" with a CIE coordinate of (0.684, 0.301). The optimized parameters are $L$=235 nm, $W$=85 nm, $\theta$=47.9º, $\Lambda_x$=436 nm, $\Lambda_y$=432 nm, and $H_{Si}$=360 nm.

The parameters of the antenna design are represented as a one-dimensional vector $P = [L, W, \theta, \Lambda_x, \Lambda_y, H_{Si}]$, where these geometric parameters are shown in Fig. 1c. In addition, $\Lambda_x$ and $\Lambda_y$ denote the pitch along *x* and *y* directions, respectively. We manually generated the initial set of parameters to bring it within the red region of the CIE map. Based on this initial starting input parameters, FDTD simulations were then carried out to simulate the optical field distribution and reflectance spectrum, which was used to calculate the coordinate (*x*, *y*) in the chromaticity diagram.

In addition, the cost function $F$ will be chosen between the 1$^{st}$ cost function "*D*" and the 2$^{nd}$ cost function "*x*". In the first step, we set "*F=D*" to minimize "*D*" for highly saturated red. After that, its first-order derivative $\frac{\partial F}{\partial P_i}$ was numerically calculated for each variable $P_i$ based on $\frac{F(P_i+\Delta P_i)-F(P_i)}{\Delta P_i}$. Finally, we obtain the optimized parameters for the minimum value of "*F*" based on the gradient descent approach after a number of iterations, *i.e.*, 60 steps, to get a convergent result. Once the optimization is stuck at a local minimum, the cost function "*F*" will be switched to "*F=x*", in the next round of optimization for another 20 iterations, until reaching another local minimum. The switching process of the cost function continues, until the optimization result does not change anymore. A more detailed explanation of this algorithm flow is shown in the Supplementary Fig. 4. Based on our experience, HSQ resist remains on the silicon nanostructures after fabrication with a thickness of ~60 nm, which was kept constant in calculations throughout the manuscript.

Next, we performed the search for the optimized geometry under both *x*- and *y*-polarized incident conditions. First, we start our analysis with the q-BIC resonance excited along *y*-polarization. To our knowledge, these *y*-polarized q-BIC resonances have not been studied so far for the nanoellipse array. In the optimization process, the initial starting point of the geometrical parameters are *L*=270 nm, *W*=100 nm, θ=65º, $\Lambda_x$=460 nm, $\Lambda_y$=420 nm, and $H_{Si}$=350 nm. Fig. 2c presents the corresponding optimization trajectory in the CIE chromaticity diagram after 120 iterations, where the optimized parameters of Si nanostructures achieving the highly saturated red are *L*=235 nm, *W*=85 nm, *θ*=47.9º, $\Lambda_x$=436 nm, $\Lambda_y$=432 nm, and $H_{Si}$=360 nm. The corresponding CIE 1931 chromaticity diagram is shown in Fig. 2c with a CIE coordinate of (0.684, 0.301), where the reflectance spectrum shown in Fig. 1b with two prominent q-BIC peaks at red wavelengths. Here, these two red q-BICs do not have the high-order diffraction since these red wavelengths are above the wavelength criterion $\lambda_0$, which can be calculated by:

$$\lambda_0 = n_{substrate} \times \max\{\Lambda_x, \Lambda_y\}. \tag{1}$$

In addition, Supplementary Fig. 5 presents the transmittance and absorptance spectra, where Supplementary Fig. 6 presents the evolution of reflectance spectra with respect to different tilt angle *θ* of the nano ellipses. It shows that the reflection peaks due to the q-BIC resonance will slowly disappear when *θ* is changed from 47.9º to 0º. At the same time, the corresponding *Q*-factor will be increased as shown in Supplementary Fig. 7. Similarly, the optimization for *x*-polarized incidence condition is shown in the Supplementary Figs. 8-9 to achieve a saturated red with a CIE coordinate of (0.678, 0.307), but with a lower reflectance as compared to the *y*-polarized design.

**Experimental fabrication and characterization.** Experimentally, we choose the design as optimized under *y*-polarized condition, due to its higher reflectance and better color saturation as compared to the *x*-polarized case. We fabricated the Si nanostructures by using high-resolution electron-beam lithography (EBL) and followed by reactive-ion etching (see details in Methods). Briefly, amorphous Si layer was grown onto the quartz substrate by using plasma-enhanced chemical vapor deposition (PECVD) method. After that, 100-nm-thick hydrogen silsesquioxane (HSQ) resist (Dow Corning, XR-1541-006, 6% dissolved in methyl isobutyl ketone) was spin coated onto the sample surface at 5k rpm, followed by EBL and dry etching processes.[57] The detailed fabrication process is shown in Supplementary Fig. 10 with the corresponding process parameters described in the Methods section. Fig. 3 presents the corresponding nanofabrication and characterization results.

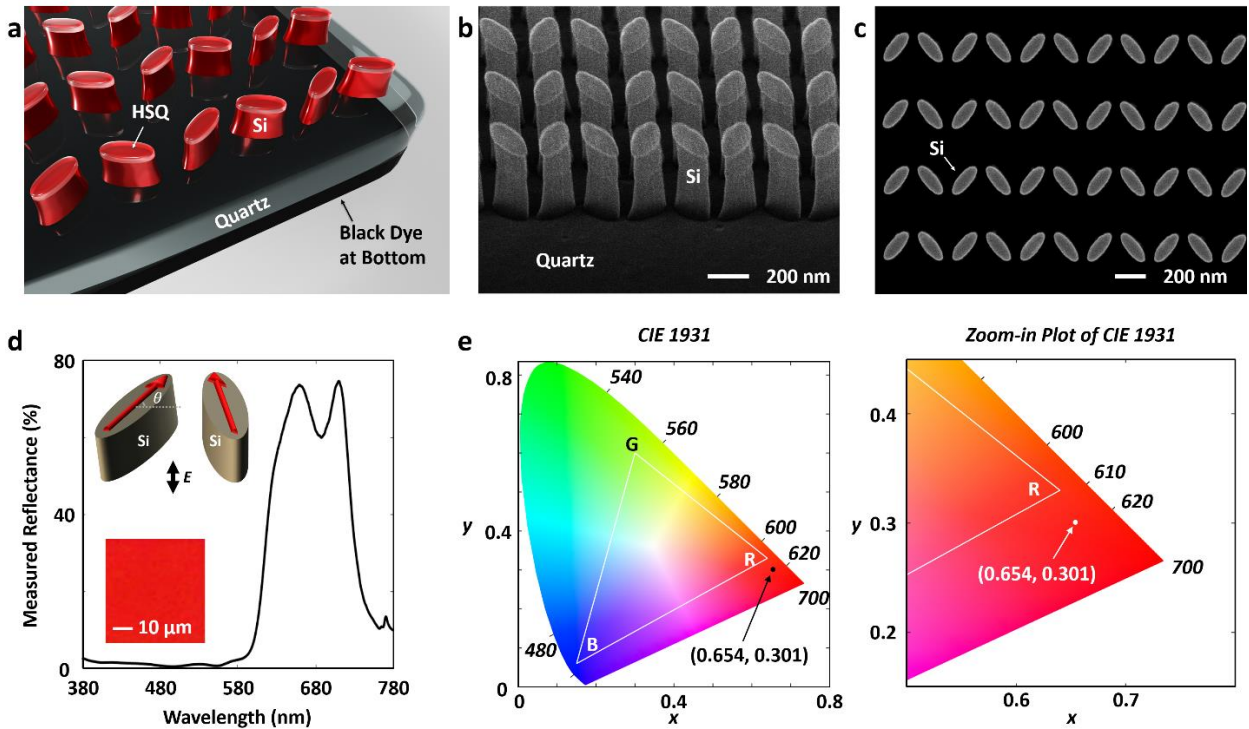

**Fig. 3 | Highly saturated red pigment under *y*-polarized incidence condition with the black dye on the back side of the quartz substrate. a,** Schematic illustration on the red pixel with q-BIC resonances. The black dye at the back side of the quartz absorbs the undesired reflection from the bottom air-quartz interface of the substrate. **b-c,** Scanning electron microscope (SEM)

images of the tilted side view and top view of the fabricated Si nanoantenna array with the highly saturated red. **d,** Measured reflectance spectra for the highly saturated red under *y*-polarization with the black dye at the back side of quartz substrate. The inset image is the optical microscope image as captured by ×5 objective lens. **e,** CIE 1931 chromaticity diagram for the measured reflectance spectrum in **d**.

To minimize the undesired reflectance from the bottom quartz-air interface, we coat it with an absorbing pigment analogous to Morpho wing scales that incorporate melanin below the multilayered ridges[58, 59] to absorb stray light, which would otherwise scatter back and significantly reduce color saturation.[59] In our experiment, black ink was coated onto the back side of the quartz substrate as shown in the corresponding schematic in Fig. 3a. The deposition of black dye is carried out by scribing "Sharpie Permanent Marker (model number: MK-SP-FINE-BLK)" onto the back side of quartz substrate uniformly. In addition, Fig. 3b and Fig. 3c present the side view and top view of the scanning electron micrograph (SEM) images of the fabricated Si nano ellipse antenna array.

Fig. 3d presents the measured reflectance spectrum and the inset figure is its optical micrograph of the final sample (see Methods section). This reflectance spectrum was measured by a micro-spectrophotometer (CRAIC UV-VIS-NIR QDI 2010, ×5 objective lens, NA=0.12) and the corresponding CIE coordinate was then calculated based on the reflectance spectrum.[15] This red saturation is beyond the sRGB triangle in the CIE chromaticity diagram as shown in Fig. 4e, with the corresponding CIE coordinate of (0.654, 0.301). In comparison, Supplementary Fig. 11 presents the measured reflectance spectrum for the same red pixel without the black dye at the back side of the quartz substrate, where the corresponding CIE coordinate is only (0.613, 0.301).

It shows that this additional black dye on the back side of the quartz substrate is able to improve the red saturation, because it is able to reduce the overall background signal in blue and green color regimes of the reflectance spectrum.

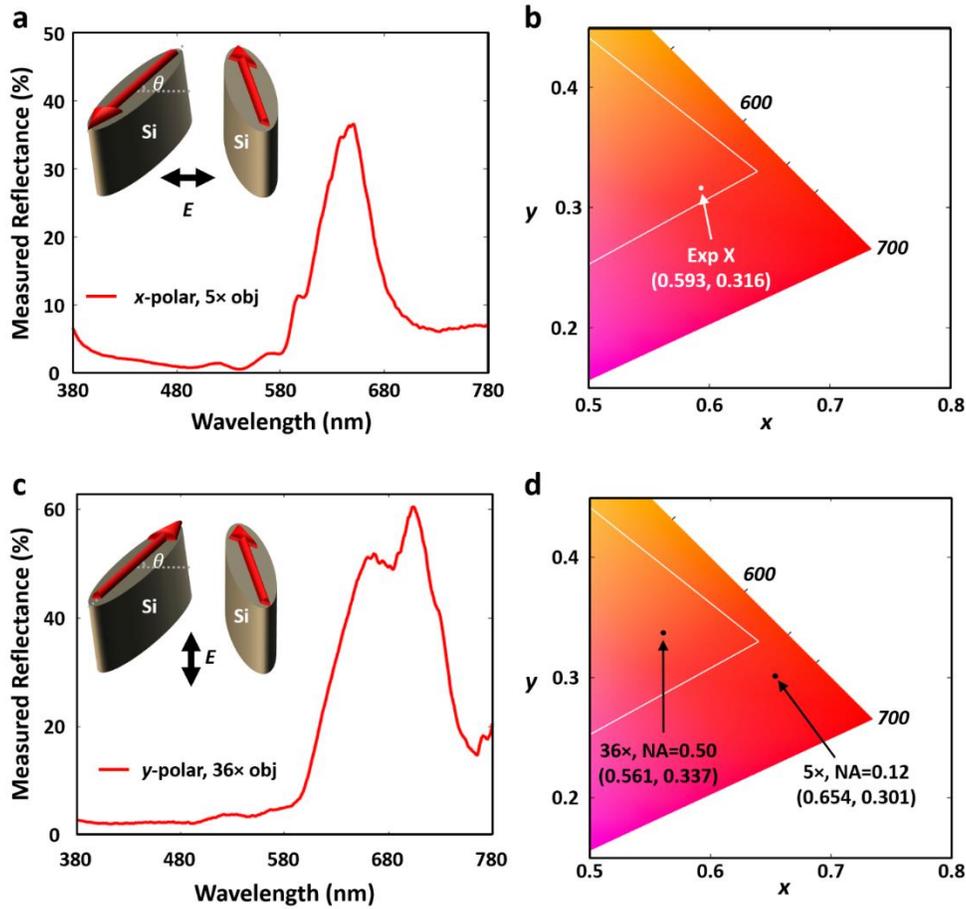

**Fig. 4 | Characterization of the fabricated red pixel with different polarization conditions and numerical apertures of the objective lens. a-b,** Measured reflectance spectrum under *x*-polarized incidence condition and the corresponding CIE coordinate a ×5 objective lens with a NA of 0.12. **c-d,** Measured reflectance spectrum at *y*-polarized incidence conditions with a ×36 objective lens with a NA of 0.50.

We also investigate the influence of incidence polarization, numerical aperture and viewing angle on the achieved color saturation. Fig. 4a presents the measured reflectance spectra for the same red pixel as fabricated in Fig. 3, by using the ×5 objective lens (NA=0.12) under *x*-polarized incidence condition. The corresponding CIE coordinate shifts to (0.593, 0.316) as shown in Fig. 4b, where the maximum reflectance is reduced to ~36%. This measurement demonstrates that our red pixel based on q-BIC resonance is strongly dependent on the incident light polarization. In addition, Figs. 4c-d present the investigation on the influence of numerical aperture on the achieved color saturation, under *y*-polarized incidence condition. It shows a reduced color saturation with increasing numerical aperture, with a shift in corresponding CIE coordinates in Fig. 4d. Thus, the most saturated reds are seen within a narrow collection angle and illumination angle (*i.e.*, a solid angle Ω of 6.9°), corresponding to an NA of 0.12. The dependences of polarization and incident angles are also evident from the angle resolved reflectance measurements at back focal plane as shown in Supplementary Fig. 12. Furthermore, due to the antenna array effect, the designed BIC antenna with the ultrahigh red saturation in this paper is iridescent. Its color will be changed slightly when the view angle is changed, where the corresponding detailed comparisons are shown in Supplementary Fig. 13 with the tilt angle of 0 degree and ±15 degree. Therefore, these characterization results in Fig. 3 and Fig. 4 show that ultra-highly saturated reds are observed under specific conditions, *i.e. y*-polarized illumination, the narrow angle of incidence and zero tilt angle.

In our design, we manage to suppress the high-order q-BIC modes at blue and green wavelengths by exploring the following two effects, *i.e.*, intrinsic absorption coefficient *k* of amorphous silicon and the substrate-induced leakage effect, where the detailed simulation comparisons are shown in Supplementary Fig. 14. For instance, Supplementary Fig. 14b shows that high-order q-BIC modes will appear if the intrinsic absorption coefficient *k* of amorphous silicon is set to zero, since the q-BIC resonances are very sensitive to *k*. Moreover, Supplementary

Figs. 14c-d present the investigation when the quartz substrate is absent, where many high-order q-BIC modes are emergent. It is because that the presence of a high-index substrate will break the out-of-plane symmetry of the optical system and open up high-order diffraction channels for these q-BIC modes.[60] In this way, the at-Γ BICs protected by the in-plane symmetry will be destroyed and converted to leaky resonances.

We applied the following protocol to characterize and calculate the achieved CIE 1931 coordinates based on the measured reflectance or transmittance spectra. The approach could be useful in standardizing future work for objective benchmarking. First, the calculation of CIE 1931 coordinate requires the full spectral information from 380 nm to 780 nm (anything less would artificially boost saturation). To simplify matters and avoid the influence of the light source, the sample is assumed to be illuminated by a perfect broadband light source with unity relative power from 380 nm to 780 nm. Then, based on this full spectral information, the CIE coordinates of the color pixels could be accurately calculated based on the equations S1-S5 in Supplementary Fig. 15.

**Conclusions**

In this paper, we present a Si antenna design with quasi bound-state-in-the-continuum resonances to achieve highly saturated red, which is beyond the sRGB triangle in the chromaticity diagram. The algorithm used aided the discovery of new *y*-polarized quasi-BIC modes. These fundamental and higher-order quasi-BIC modes closely placed within the spectral passband of the ideal Schrödinger red provide the desirable sharp transition necessary for Schrödinger color saturation. To the best of our knowledge, the results represent the highest red saturation achieved by nanostructures with full spectral characterization across from 380-780 nm. Notably, due to the angle and polarization dependent nature of the color, the presented design could be exploited for unique visual effects that are absent in pigments. Moreover, different configurations based on single asymmetric dielectric nanostructures could be explored further with smaller foot prints towards higher printing resolution.[35, 61]

**Methods**

**Numerical Simulations.** A commercial software (Lumerical FDTD Solutions) was used to do the Finite-difference time-domain (FDTD) simulations. Along $x$-/$y$-directions, periodic boundary conditions were used and perfect matched layer (PML) was used along $z$-direction. The $n$ and $k$ values of the amorphous Si were measured by ellipsometer as shown in Supplementary Fig. 1. Based on the Lumerical script and achieved reflectance spectra by simulation, the CIE 1931 chromaticity diagram is calculated by the color matching functions, where the detailed formulation could be found in Supplementary Fig. 15.

**CVD Growth of Si Film, Nanofabrication and Si Etching.** Plasma-enhanced chemical vapor deposition (PECVD) method was used to grow the amorphous Si film onto quartz substrate. The etching masks were fabricated based on 100-nm-thick HSQ, which was obtained by spin coating the resist (Dow Corning XR-1541-006) at 5k round-per-minute (rpm). The layout file of nano ellipse array was created by using the software Auto CAD, and then this file was converted into the exposure file by another software Beamer (GenISys GmbH). The electron beam lithography (Elionix) was then conducted with an acceleration voltage of 100 keV, beam current of 200 pA, and an exposure dose of ~12 mC/cm$^2$. The sample was then developed by a NaOH/NaCl salty solution (1% wt./4% wt. in de-ionized water) for 60 seconds and immersed in de-ionized water.[62]

Next, the sample was immediately rinsed by acetone, isopropanol alcohol (IPA) and dried by a continuous flow of nitrogen. Si etching was then carried out by using inductively-coupled-plasma (ICP, Oxford Instruments Plasma Lab System 100), with $Cl_2$ gas chemistry at 40 °C.

**Characterizations.** The optical reflectance spectrum was measured by using a CRAIC micro-spectrophotometer QDI 2010 (×5 objective lens, Zeiss A-plan with a numerical aperture of 0.12), under the illumination of polarized broadband light source (75 W Xenon Lamp). The absolute reflectance spectra were obtained, where we calibrated our reflectance measurements on a certified calibration aluminum sample from CRAIC Technologies. Moreover, for the acquisition of the optical microscope image in Fig. 3, a white color balancing was first carried out on the calibration aluminum sample. The images were then taken by using a microscope (Olympus MX61) with the "analySIS" software, a ×5 objective lens (MPlanFL N, NA=0.15), a camera (Olympus SC30) with an integration time of 20 ms, halogen light source (U LH100 3) with a linear polarizer. The SEM images were taken with an electron acceleration voltage of 10 keV (SEM, Elionix).

**Online content**

Any methods, additional references, Nature Research reporting summaries, source data, extended data, supplementary information, acknowledgements, peer review information; details of author contributions and competing interests; and statements of data availability are available at the website.

**Data availability**

The data that support the figures and other findings of this study are available from the corresponding authors upon reasonable request.

**Acknowledgements**

This work is supported by Agency for Science, Technology and Research (A*STAR) career development award (CDA) with the grant number of 202D8088. In addition, Z.D. and J.K.W.Y. would like to acknowledge the funding support from A*STAR AME IRG with the project number of A20E5c0093. C.W.Q. is also supported by the grant (R-261-518-004-720) from Advanced Research and Technology Innovation Centre (ARTIC).

**Author contributions**

Z.D., L.J, C.-W.Q. and J.K.W.Y. conceived the concept of using nanostructures with quasi-BIC resonances to achieve highly saturated reds. Z.D. and J.K.W.Y. designed the experiments, fabricated, characterized the samples, and wrote the manuscript. L.J. and C.-W.Q. performed the optimized algorithm. S.D.R. performed the FDTD simulations. H.W. and Z.D. performed the optical characterizations. H.W. performed the CIE coordinate calculations. Y.C. performed the theoretical analysis on BIC. F.T. and Z.D. performed the dry etching process. J.F.H. performed the preliminary finite-difference time-domain simulations. S.G. performed the back focal plane measurements. R.J.H.N. and Q.R. participates in the discussions. All authors analyzed the data, read and corrected the manuscript before the submission. Z.D and L.J. contributed equally to this work.

**Competing interests**

The authors declare that they have no competing interests.

**Additional information**

**Correspondence and requests for materials** should be addressed to J.K.W.Y., C.-W.Q. or Z.D.

*Supplementary Information for:*

# Schrodinger's Red Pixel by Quasi Bound-State-In-Continuum

Zhaogang Dong[1,†,*], Lei Jin[2,3,†], Soroosh Daqiqeh Rezaei[4], Hao Wang[4], Yang Chen[2], Febiana Tjiptoharsono[1], Jinfa Ho[1], Sergey Gorelik,[5] Ray Jia Hong Ng[4], Qifeng Ruan[4], Cheng-Wei Qiu[2,*], and Joel K. W. Yang[1,4,*]

[1]Institute of Materials Research and Engineering, A*STAR (Agency for Science, Technology and Research), 2 Fusionopolis Way, #08-03 Innovis, 138634 Singapore

[2]Department of Electrical and Computer Engineering, National University of Singapore, 4 Engineering Drive 3, Singapore 117583, Singapore

[3]College of Electronic and Information Engineering, Hangzhou Dianzi University, Hangzhou, China, 310018

[4]Singapore University of Technology and Design, 8 Somapah Road, 487372, Singapore

[5]Singapore Institute of Food and Biotechnology Innovation, A*STAR (Agency for Science, Technology and Research), 31 Biopolis Way, #01-02 Nanos, Singapore 138669

*Corresponding author. Email: joel_yang@sutd.edu.sg, chengwei.qiu@nus.edu.sg, dongz@imre.a-star.edu.sg.

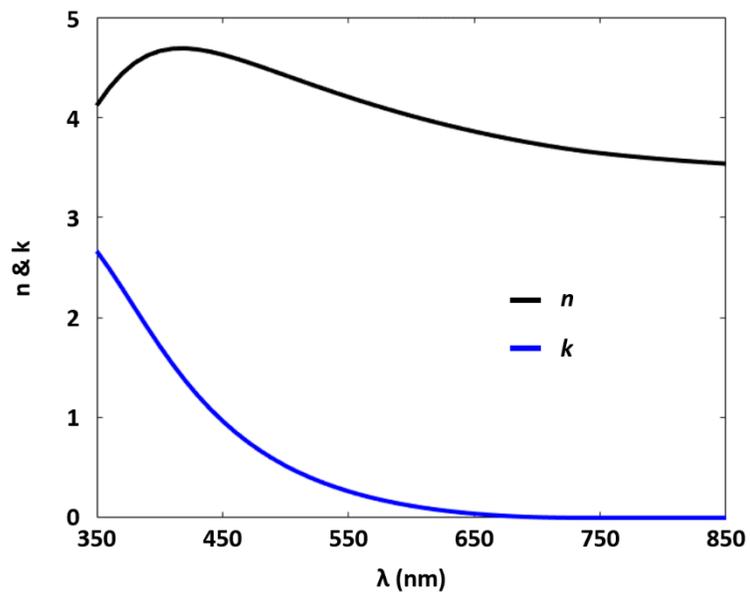

**Supplementary Fig. 1 | Measured *n* and *k* values for the amorphous Si films as grown by inductively coupled plasma (ICP) chemical vapor decomposition (CVD).**

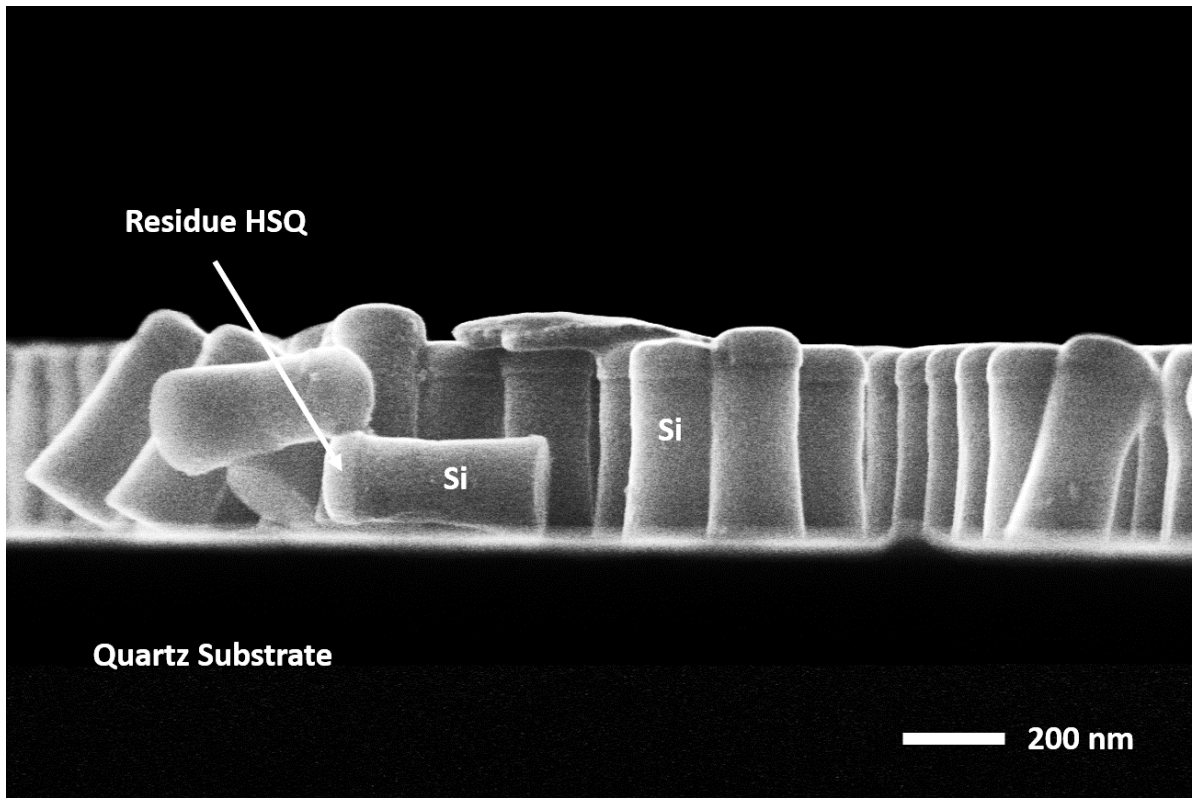

**Supplementary Fig. 2 | Cross sectional profile for Si nanostructures as etched by ICP with Cl₂ gas chemistry.** The dry etching of Si was carried out by using inductively-coupled-plasma (ICP, Oxford Instruments Plasmalab System 100), based on Cl₂ gas chemistry with a gas flow rate is 22 sccm at 40 °C. The etched Si nanostructures are not having the perfect 90 degree cross sectional profile. Instead, it has two portions. First, the top ~150 nm segment (*i.e.*, the part close to the HSQ resist) has as ~90° side wall profile, while the bottom segment tapers outwards to a trapezoidal shape with side wall angle of ~82°.

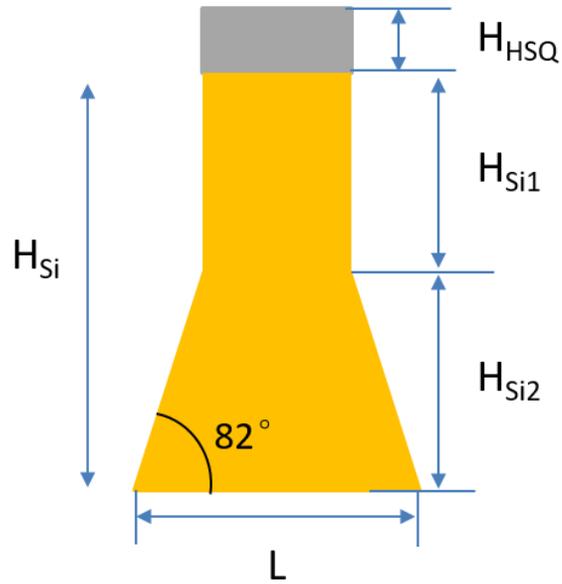

**Supplementary Fig. 3 | Cross sectional profile of Si nanostructures.** The total height of the Si antenna is denoted as $H_{Si}$, which consists of two segments. $H_{Si1}$ denotes the height of the top 150 nm segment (*i.e.*, the part close to the HSQ resist) with a 90º side wall profile. The height of the bottom segment is denoted as $H_{Si2}$ with a taper being modelled a trapezoidal shape with side wall angle of ~82º.

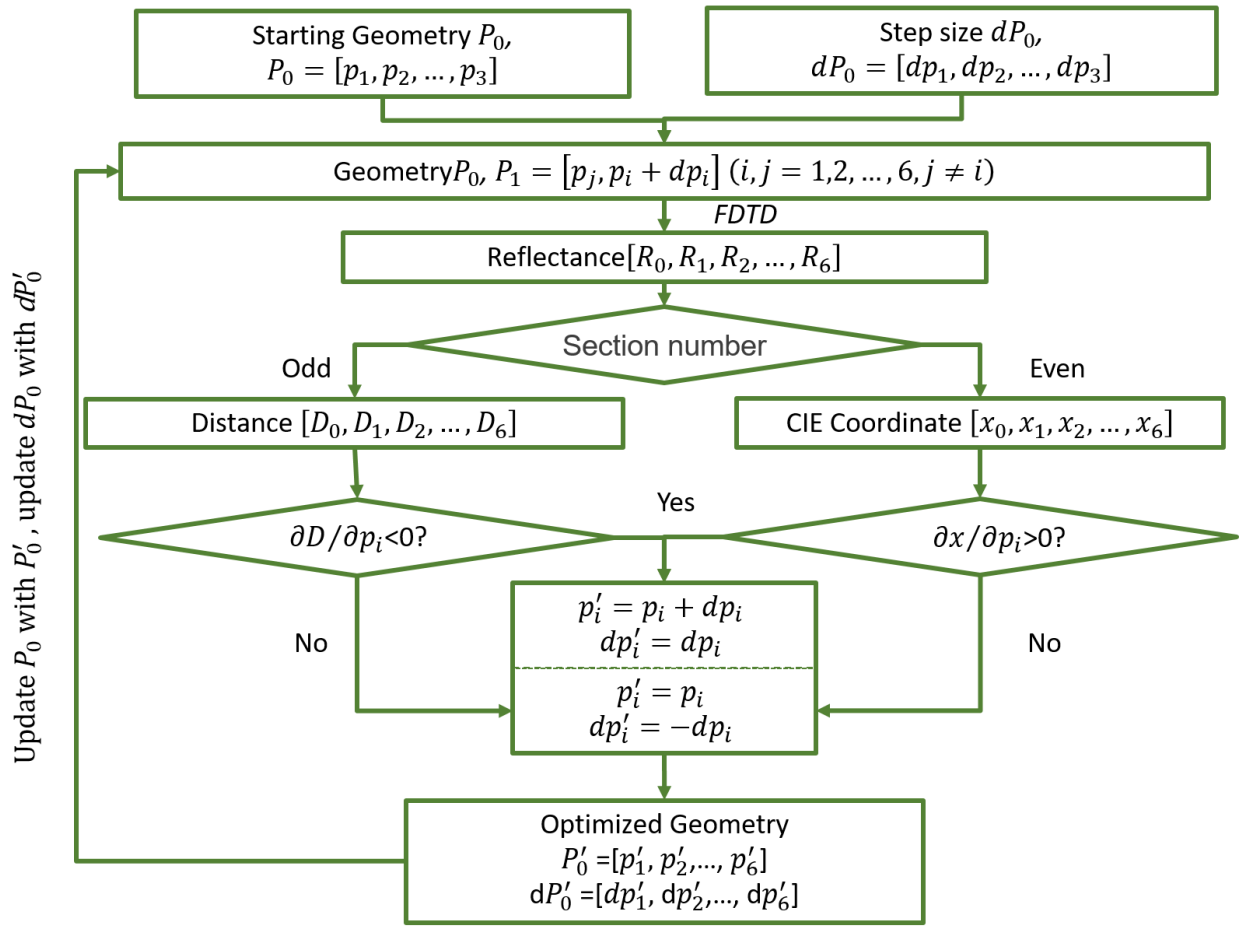

**Supplementary Fig. 4 | Detailed flow of the gradient descent algorithm.**

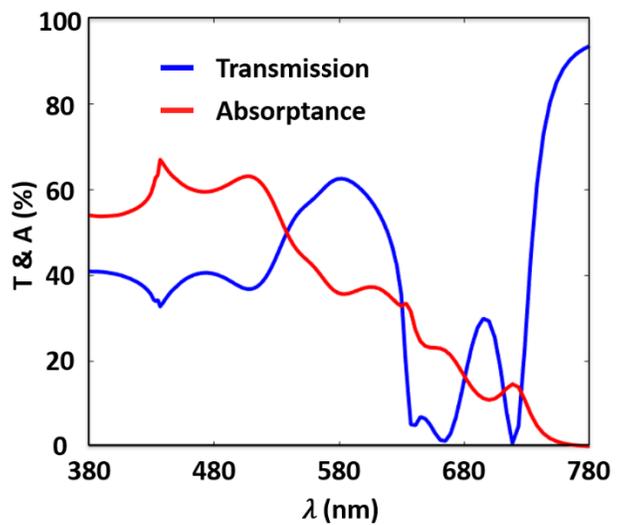

**Supplementary Fig. 5 | Simulated transmission (T) and absorptance spectra under *y*-polarized condition**.

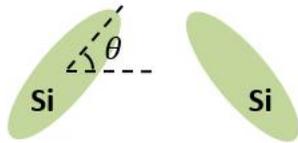
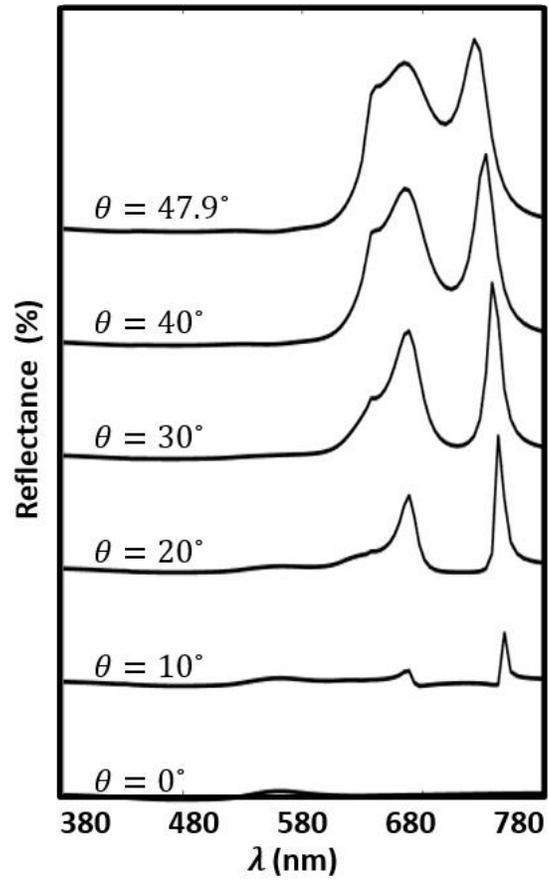

**Supplementary Fig. 6 | Evolution of the simulation reflectance spectra with respect to different tilt angle θ at *y*-polarized incidence condition.**

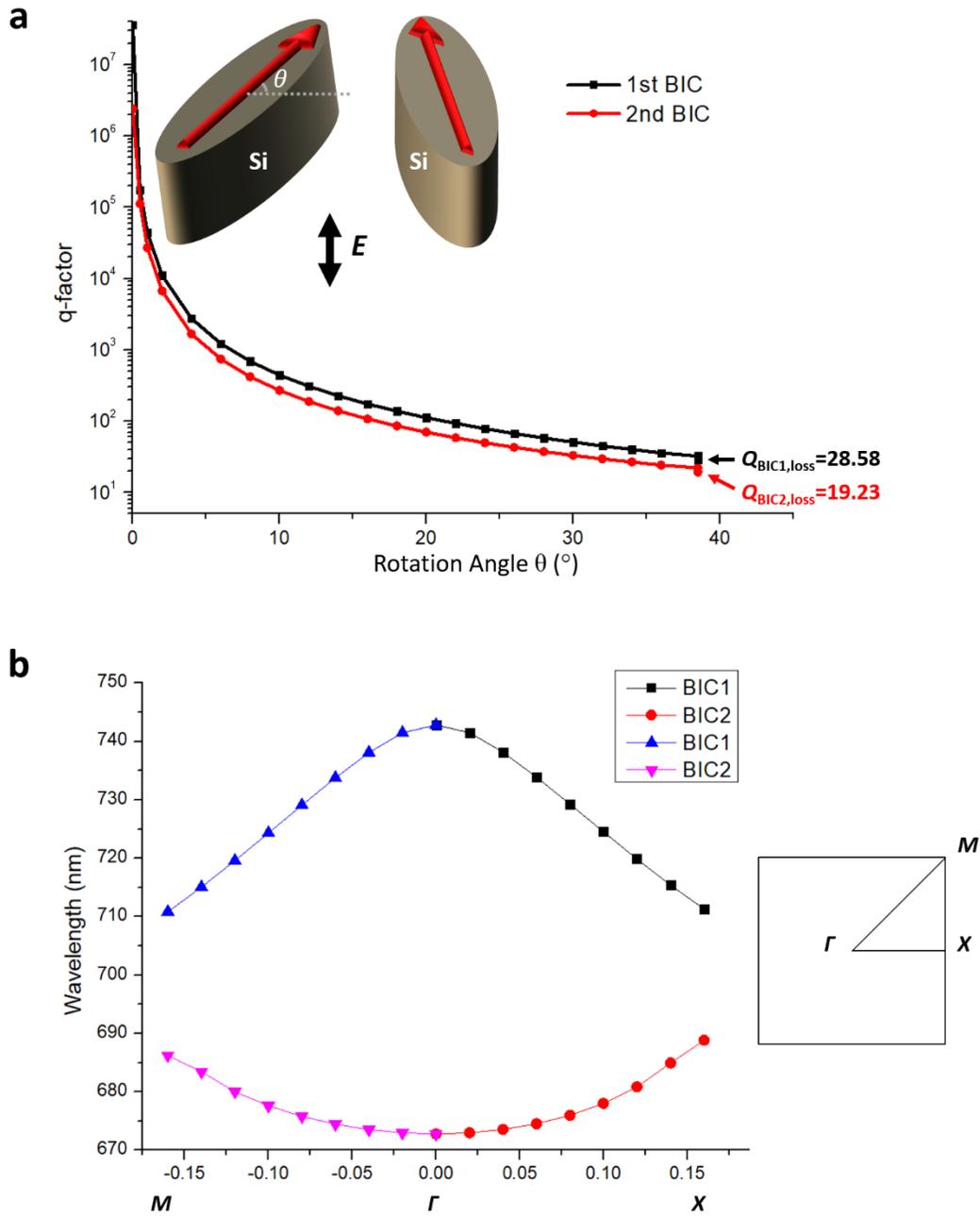

**Supplementary Fig. 7 | Q-factor and mode analysis. a**, Dependence of the antenna's *Q*-factor for 1st BIC mode and 2nd BIC mode, with respect to different tilt angle θ at *y*-polarized incidence condition. This simulation was carried out by setting the material loss to be zero. The *Q* factors for both BIC modes with real material loss at the tilt angle θ of 38° are shown for benchmarking purpose. **b**, BIC resonance wavelengths in the *k*-space.

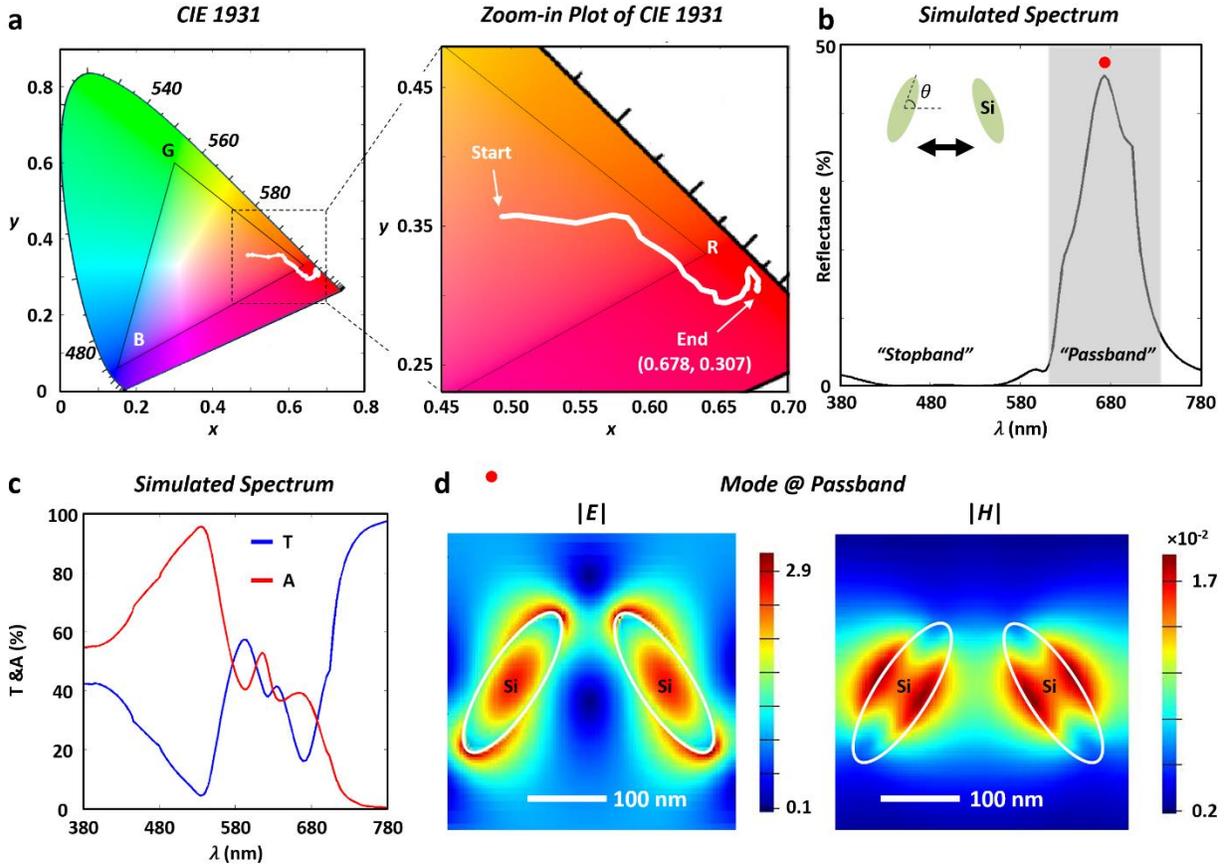

**Supplementary Fig. 8 | Optimized Si nanostructures by gradient descent optimization algorithm under *x*-polarized incidence condition**. **a,** Trajectory of the CIE coordinates during the optimization process from "start" to "end" with a CIE coordinate of (0.678, 0.307). The initial starting point of the geometrical parameters are $L$=200 nm, $W$=100 nm, $\theta$=65°, $\Lambda_x$=600 nm, $\Lambda_y$=400 nm, and $H_{Si}$=300 nm. After 160 iterations, the optimized parameters are $L$=260 nm, $W$=85 nm, $\theta$=56.7°, $\Lambda_x$=394 nm, $\Lambda_y$=481 nm and $H_{Si}$=397 nm. **b-c,** Simulated reflection, transmission and absorption spectra based on the optimized nanoantennas. **d,** Mode patterns of the electric field ($|E|$) and magnetic field ($|H|$) at the reflectance peak wavelength of the passband with a q-BIC resonance. This mode which is a q-BIC mode, as it is dependent on the tilt angle $\theta$. When the tilt angle $\theta$ is approaching to 90 degrees, this quasi BIC will slowly disappear as shown in Supplementary Fig. 9.

S9

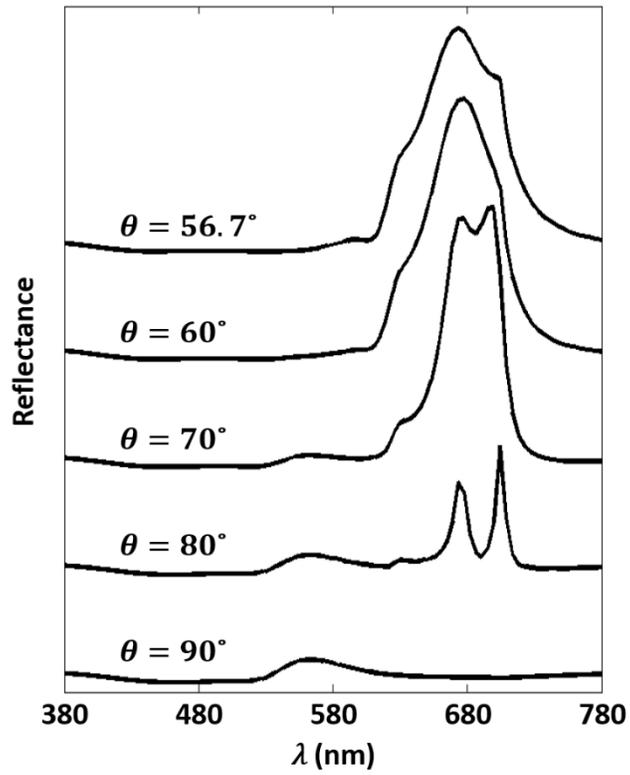

**Supplementary Fig. 9 | Evolution of the simulation reflectance spectra with respect to different tilt angle θ at *x*-polarized incidence condition.**

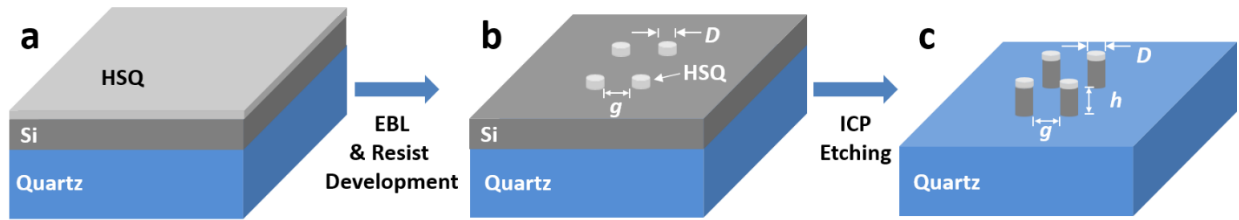

**Supplementary Fig. 10 | Fabrication process of silicon nano ellipses on quartz substrate. a,** Hydrogen silsesquioxane (HSQ) resist spin coated onto the silicon substrate as grown on quartz substrate. **b,** E-beam exposure to fabricate the HSQ resist masks. **c,** Inductively-coupled plasma (ICP) for the reactive etching of silicon with $Cl_2$ gases at 40 °C.

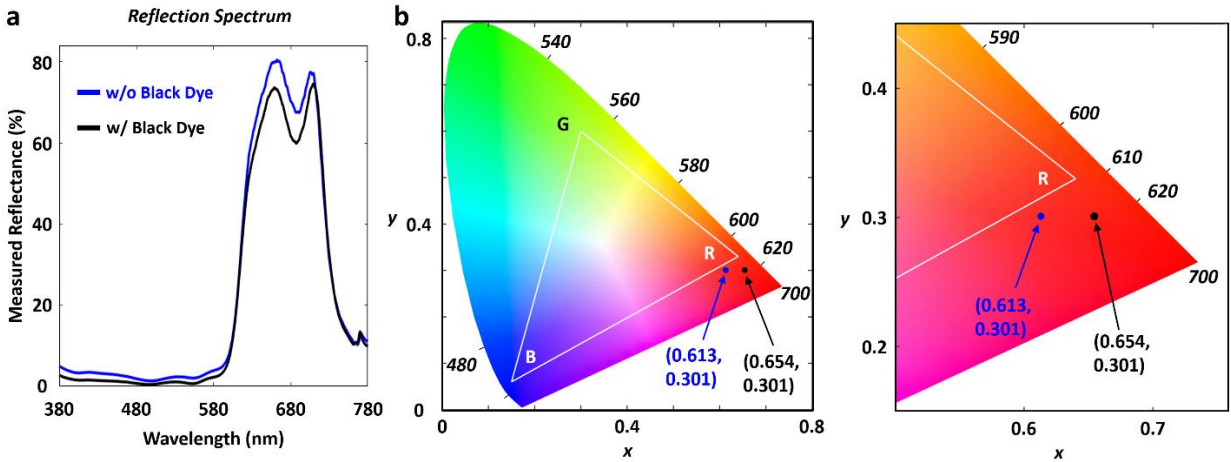

**Supplementary Fig. 11 | Highly saturated red color pixel without and with the black dye on the back side of the quartz substrate under *y*-polarized incidence condition. a,** Reflectance spectra without and with black dye on the back side of the quartz substrate. **b,** The corresponding CIE coordinates for the reflectance spectra without dye (denoted by dots in blue color) and with dye (denoted by dots in black color) on the back side of the quartz substrate. It shows that the CIE coordinate is shifted from (0.613, 0.301) to (0.654, 0.301) due to the deposition of the black dye on the back side of the quartz substrate.

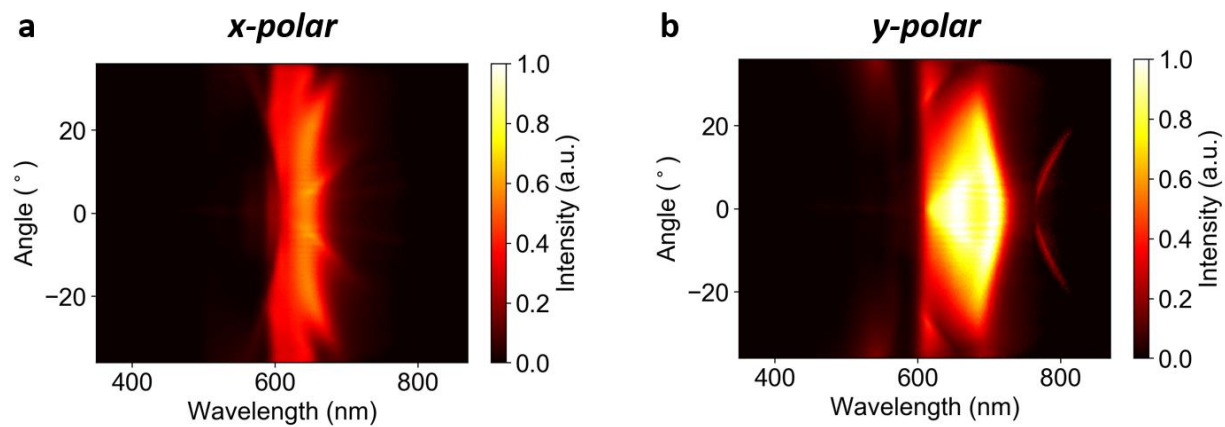

**Supplementary Fig. 12 | Relative reflectance measurements at the back focal plane with respect to an aluminum mirror. a,** *x*-polarized incidence condition. **b,** *y*-polarized incidence condition. Angle-resolved reflectance spectra were measured at the back focal plane spectroscopy. An inverted optical microscope (Nikon Ti-U) was coupled to a spectrograph (Andor SR-303i) equipped with an EMCCD detector (Andor Newton). During the reflectance measurements, light from a halogen lamp passing through a microscope field aperture was focused onto the sample surface using an objective lens (50×, NA of 0.6), where the reflectance from an aluminum mirror was used as a reference.

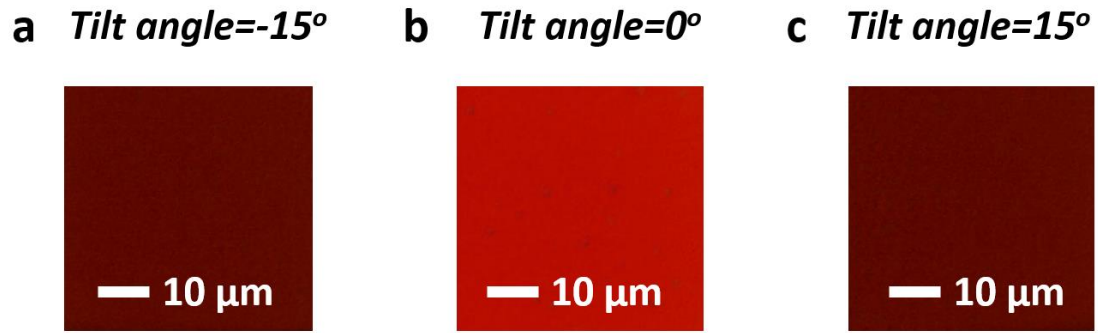

**Supplementary Fig. 13 | Red color pixel observed with the tilt angles. a,** -15 degree. **b,** 0 degree**. c,** 15 degree. The objective lens used here is ×20 objective lens (NA = 0.45). It shows that the red color pixel will change slightly when the sample is tilted.

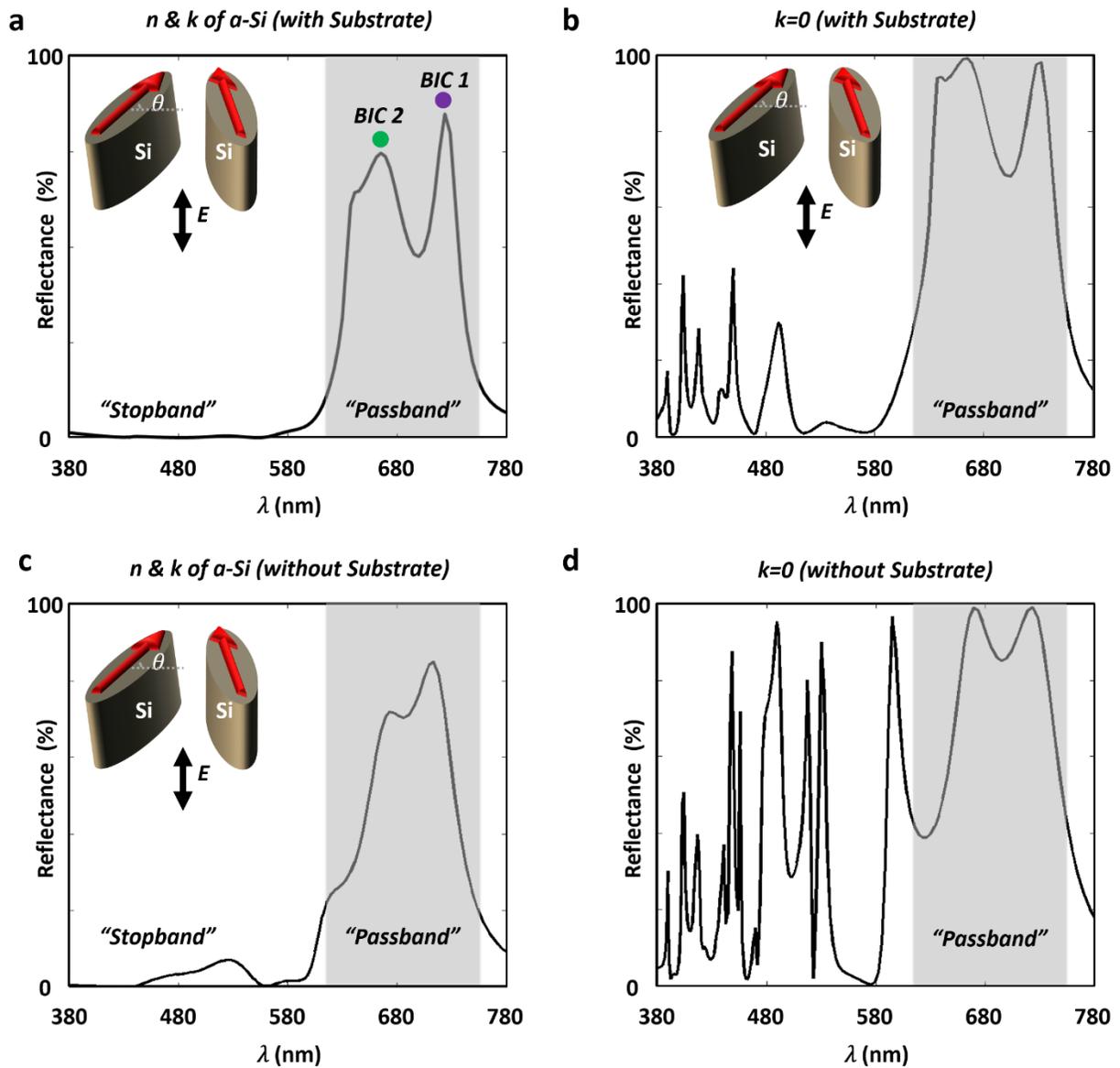

**Supplementary Fig. 14 | Simulation results on the effects of absorption extinction *k* and substrate. a,** Real *n* and *k* values of amorphous Si with quartz substrate. **b,** Setting *k*=0 for amorphous Si with quartz substrate. **c,** Real *n* and *k* values of amorphous Si without quartz substrate. **d,** Setting *k*=0 for amorphous Si without quartz substrate.

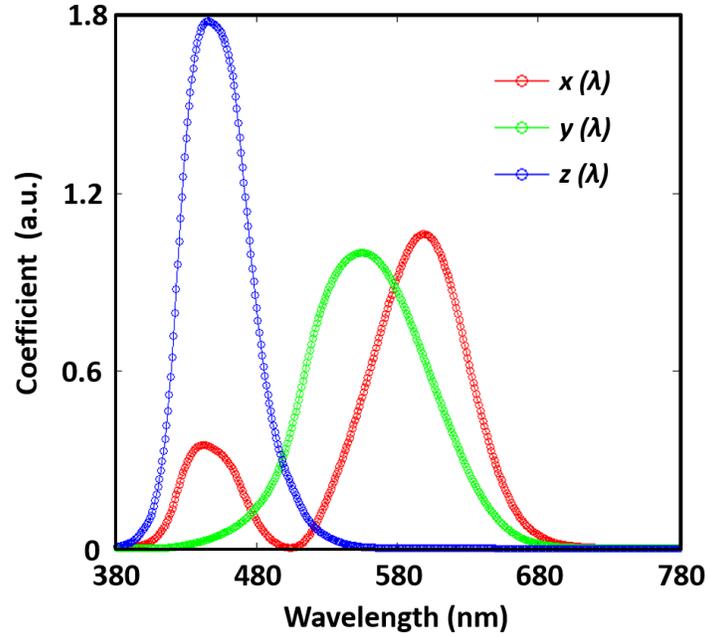

**Supplementary Fig. 15 | Coefficients of the color matching functions for "Blue", "Green" and "Red" colors across the spectrum from 380 nm to 780 nm.**[1] These color matching functions are denoted as $x(\lambda)$, $y(\lambda)$ and $z(\lambda)$, respectively. Here, $R(\lambda)$ denote the reflectance or transmittance spectrum. To obtain the CIE coordinates, the tristimulus $X$, $Y$ and $Z$ values were calculated by:[1]

$$X = \int_{380}^{780} R(\lambda)x(\lambda)d\lambda, \tag{S1}$$

$$Y = \int_{380}^{780} R(\lambda)y(\lambda)d\lambda, \tag{S2}$$

$$Z = \int_{380}^{780} R(\lambda)z(\lambda)d\lambda. \tag{S3}$$

The $X$ and $Y$ values were then normalized to obtain the CIE ($x,y$) coordinates:

$$x = \frac{X}{X+Y+Z}, \tag{S4}$$

$$y = \frac{Y}{X+Y+Z}. \tag{S5}$$

To avoid the influence of the light source, the sample is assumed to be illuminated by a perfect broadband light source from 380 nm to 780 nm.